\documentclass[a4paper,aps,prb,twocolumn, superscriptaddress]{revtex4-1}
\usepackage{graphicx}
\usepackage{dcolumn}
\usepackage{bm}
\usepackage{amsmath}
\usepackage{amsfonts}
\usepackage{amssymb}%
\usepackage{here}
\usepackage{color}

\usepackage{txfonts}

\begin{document}
%
%
% Title
%
%
\title{Effects of Cavitation on K\'arm\'an Vortex Behind Circular-Cylinder Arrays}
\author{Yuta Asano}
\email{yuta.asano@issp.u-tokyo.ac.jp}
\affiliation{Institute for Solid State Physics, The University of Tokyo, Kashiwa, Chiba 277-8581, Japan}
\author{Hiroshi Watanabe}
\affiliation{Institute for Solid State Physics, The University of Tokyo, Kashiwa, Chiba 277-8581, Japan}
\affiliation{Department of Applied Physics and Physico-Informatics, Keio University, Yokohama, Kanagawa 223-8522, Japan}
\author{Hiroshi Noguchi}
\affiliation{Institute for Solid State Physics, The University of Tokyo, Kashiwa, Chiba 277-8581, Japan}
%
%
% Abstract
%
%
\begin{abstract}
The effects of cavitation on the flow around a circular-cylinder array are studied by using a molecular dynamics simulation. 
Cavitation significantly affects on vortex shedding characteristics.
As the cavitation develops, the vibration acting on the cylinders decreases and eventually disappears. The further cavitation development generates a longer vapor region next to the cylinders, and the vortex streets are formed at further positions from the cylinders.
The neighboring  K\'arm\'an vortexes are synchronized in the antiphase in the absence of the cavitation. 
This synchronization is weakened by the cavitation, and an asymmetric wake mode can be induced.
These findings help mechanical designs of fluid machinery that include cylinder arrays.
\end{abstract}
\maketitle
%
%
% Introduction
%
%
\section{Introduction}
Cavitation is a flow phenomenon induced by local pressure changes in a liquid that causes bubble generation, growth, and collapse~\cite{Brennen95}.
Cavitation that occurs during the operation of fluid machinery, such as ship propulsion devices and turbomachinery, can result in various adverse effects, including decreased measures of performance, increased noise and vibration, and erosion caused by impact pressure from bubble collapse.
Moreover, since flow characteristics significantly change during cavitation, it is challenging to design a structure that prevents resonance caused by fluid excitation vibration.
Therefore, a current major challenge in mechanical design is to develop structures that can withstand or prevent cavitation during operation, thereby saving energy and environmental degradation.
Researchers are addressing this challenge by examining cavitation conditions and the influence of bubbles in the flow field on fluid dynamics.

Researchers have conducted experiments and computer simulations based on the Navier-Stokes equations that focus on the influence of cavitation on fluid machinery~\cite{ljt16, sap17}.
The bubble nuclei contained in liquid is an essential factor in the inception of cavitation.
Notably, the interaction between the vortex structure and the dynamics of bubble nuclei is a critical factor in a turbulent flow~\cite{Arndt02}. 
However, the mechanism of how the interaction between bubble nuclei and vortex structure in a flow field creates cavitation is still unclear.
Therefore, to discuss the mechanism of cavitation generation, it is necessary to examine the phase transition dynamics in the vortex flow directly.

The flow of Newtonian fluids around a circular cylinder is one of the most basic examples of vortex flow and has been the subject of numerous experiment~\cite{williamson96}~.
This flow includes various physics, such as vortex motion~\cite{pcl82,gerrard66}, Aeolian sound~\cite{phillips56}, and drag behavior~\cite{henderson95}, which are the critical problems of fluid mechanics.
Moreover, the vibration excited by the generation of a K\'arm\'an vortex can be easily measured to investigate the interaction between a fluid and structure~\cite{bearman11}.
The flow around a cylinder is characterized by Reynolds number $Re=\rho D V \eta$ ($\rho$: density, $D$: cylinder diameter, $V$: flow velocity, $\eta$: viscosity).
At $Re<49$, the flow is steady and twin symmetric vortices are formed behind the cylinder. Twin vortices stretch with increasing $Re$.
When $Re=49$, the extended twin vortex becomes unstable, initiating the release of the K\'arm\'an vortex.
In the range of $49\le Re \lesssim 190$, vortex shedding occurs periodically and can be described by the Strouhal number $St=fD/V$ ($f$: shedding frequency), the dimensionless shedding frequency~\cite{roshko54}:
\begin {eqnarray}
St & = & 0.212-\frac{4.5}{Re}. \label {eq0}
\end {eqnarray}
When  $Re$ is further increased, a three-dimensional vortex structure appears, and the wake of the cylinder transitions to turbulent flow.

When several objects are in close proximity, flow around one of the objects is influenced by the flows around the neighboring objects.
Flow characteristics around cylinders depend on the number and arrangement of the cylinders~\cite{za16}, such as the arrangement of electric wires, bridge girders, and tubes in cooling systems.
The most basic example of such a flow is the flow around the circular cylinders arranged in a side-by-side configuration.
In this configuration, various flow characteristics appear, such as an interference of  K\'arm\'an vortex and irregular changes in the gap flow, depending on the distance between the cylinders~\cite{ams03, azh17}.

Cavitation flow around isolated circular cylinders has been examined in experiments~\cite{vs65,fry84,mgo92,ss03,kcb17, kbc17}.
These studies analyzed the effect of bubble generation on the vortex structure and the behavior of pressure waves generated during bubble collapse.
For flows with cavitation, in addition to $Re$, the cavitation number also contributes to the characterization of the flow field.
The cavitation number is an index representing the ease of occurrence of cavitation, and the smaller the value, the more easily bubbles are generated.
Varga and Sevestyen have investigated the relationship between the cavitation number, cavity length, and Strouhal number in the critical and supercritical $Re$ range~\cite{vs65}.
They reported that with decreasing cavitation number, the cavity length increases and the vortex shedding frequency decreases.
Fry added to this understanding cavitation flow by investigating how bubble growth and vortex shedding processes depended on cavitation number~\cite{fry84}.
Fry concluded that the pressure wave induced by bubble collapse links to the vortex shedding.
Additionally, the vortex shedding frequency increases when bubbles stick behind the cylinder.
Matsudaira {\it{et al.}} found that high impulsive pressure was periodically generated behind the cylinder in conjunction with vortex shedding, and suggested a correlation between the Strouhal number and bubble collapse~\cite{mgo92}.
They also showed that when the cavitation number decreases, bubbles develop, and the frequency of occurrence of impact pressure and pressure fluctuation decrease, while the maximum value of impact pressure increases.
Sato and Saito found three types of bubble collapse patterns in the wake of a cylinder and concluded that the highest impact pressure was generated by the kind of collapse that collides with the wall surface~\cite{ss03}.
Kumar {\it{et al.}} investigated the dynamics of cavitation around a cylinder and showed that the cavity length is appropriately scaled during the bubble growth process~\cite{kcb17, kbc17}.
They also showed that the lifetime of bubbles behind the cylinder is different from that of an isolated bubble.

Computer simulations based on the Navier-Stokes equation have also been performed for the cavitation flow around a cylinder.
Seo {\it{et al.}} analyzed the cavitation flow around a cylinder in two-dimensional space at $Re = 200$ using a homogeneous equilibrium model\cite{sms08}.
Their analysis of the acoustic wave demonstrated that the pressure wave was generated during bubble collapse, resulting in a modification of the vortex shedding frequency.
Gnanaskandan and Mahesh also analyzed the cavitation flow around a cylinder, but they investigated the effect of bubbles on the flow field in detail~\cite{gm16}.
Their homogeneous medium model at $Re=200$ and $3900$ demonstrated that the vortex elongation term included in the vorticity equation plays an essential role in the decrease of the vortex shedding frequency due to bubble generation.
They suggested that cavitation suppresses the turbulence and three-dimensional instability of K\'arm\'an vortices because it reduces effective $Re$.

Previous numerical studies have discussed the influence of the growth and extinction of bubble nuclei in the liquid within the flow field.
However, the phase transition in the flow field should be considered in the case of cavitation in the absence of bubble nuclei in the liquid~\cite{wtu08}.
This presents a challenge since simulations based on fluid mechanics require bubble seed and their growth models~\cite{tw18}.
However, recent progress in computational power has enabled the analysis of the flow around a cylinder with molecular dynamics (MD) simulations~\cite{awn18, awn19}.
MD, which does not require the introduction of a phase transition model, has been used to simulate bubble generation and growth in a liquid in the absence of flows~\cite{ttm08, wii14}.
In the present study, we extend these MD simulations to include cavitation flow around a circular-cylinder array.

While cavitation flow around isolated cylinders has been analyzed in previous studies, cavitation flow around a group of bodies has received less attention.
Cavitation flow around a group of bodies is observed in many critical engineering systems, such as electronic cooling systems~\cite{kp10} and cavitation reactors for the biodiesel production~\cite{lhm16}.
In the present study, we use MD simulations to analyze the cavitation flow around infinitely arranged side-by-side circular cylinders and investigate the effect fo cavitation on the flow characteristics.
We also discuss the effects of synchronization between neighboring K\'arm\'an vortices.
This is the first trial to analyze the origin and effects of cavitation without the assumptions of a void-containing fluid and growth model that are required by the general hydrodynamic approach.

In Sec.~\ref{sec:method}, the simulation model and method are described.
The flow behaviors in the absence and presence of cavitation are presented in Sec.~\ref{sec:ncflow} and \ref{sec:cavflow}, respectively.
The discussion and summary are described in Sec.~\ref{sec:dis} and \ref{sec:sum}, respectively.
%
%
% Method
%
%
%
\begin{figure}[thbp]
\includegraphics{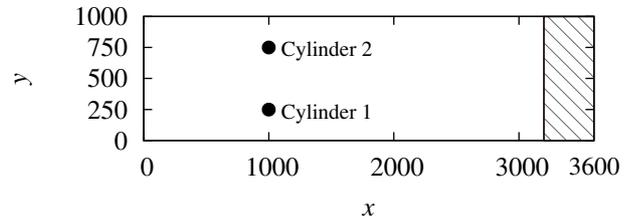}
\caption{Schematic view of the computational domain. Cylinder 1 and 2 (black circles) are located at $(x, y)=(1000, 250)$ and $(1000, 750)$, respectively, both with diameter $D=100$. The Langevin thermostat is applied in the region indicated in the shaded area ($3200\le x \le 3600$). The thickness in the depth direction is $100$.}
\label{f0}
\end{figure}

\section{Method}\label{sec:method}

Smooth-cutoff Lennard-Jones (LJ) potential $\phi$ is used to model interparticle interaction:
\begin{eqnarray}
\phi\left(r\right) &=& \left\{ \begin{array}{ll}
 \phi_{0}\left(r\right) - \phi_{0}\left(r_{\rm c}\right) + \left(r-r_{\rm c}\right) \phi_{0}'(r_{\rm c})& \left(r>r_{\rm c}\right)\\
 0 & \left(r\le r_{\rm c} \right) \end{array} \right. \label{eq1},\\
\phi_{0}\left(r\right)&=&4\epsilon \left[ \left(\frac{\sigma}{r}\right)^{12} - \left(\frac{\sigma}{r}\right)^{6} \right],
\label{eq2}
\end{eqnarray}
where $r$ is the interparticle distance, and $\epsilon$ and $\sigma$ represent the energy and length scales, respectively.
Prime represents the derivative with respect to $r$.
$r_{\rm c}$ is the cutoff distance of the potential function.
Here, we use the value of $r_{\rm c}=2.5\sigma$. % for the LJ fluid, and $r_{\rm c}=2^{1/6}\sigma$ for the Weeks--Chandler--Andersen(WCA) fluid~\cite{wca71}.
Particles whose interactions are described by Eq.~(\ref{eq1}) are called LJ particles.
In the following, all physical quantities are expressed in units of energy $\epsilon$, length $\sigma$, and time $\tau_{0}=\sqrt{m/\epsilon}$,
where $m$ is the mass of the particle.

The simulated system is a rectangular parallelepiped with dimensions $L_{x}\times L_y\times L_z=3600\times 1000\times 100$ as shown in Fig.~\ref{f0}.
The periodic boundary condition is applied to all directions.
The circular cylinders are modeled by fixing LJ particles on their surfaces.
By using the periodic boundary conditions for the flow around the two cylinders, we can simulate asymmetrical flows such as antiphase K\'arm\'an vortices and flip-flopping.
The Langevin thermostat~\cite{awn18} is used as part of the system to control the fluid inflow velocity and to equilibrate the fluid.
The friction coefficient is linearly increased from $0.0001$ to $0.1$ in the region of $3200\le x \le 3400$ and is held at $0.1$ for the remaining shaded area in Fig.~\ref{f0}.
The inflow velocity is set to $V=1$ in the $x$-direction.
\begin{figure}[htbp]
\includegraphics{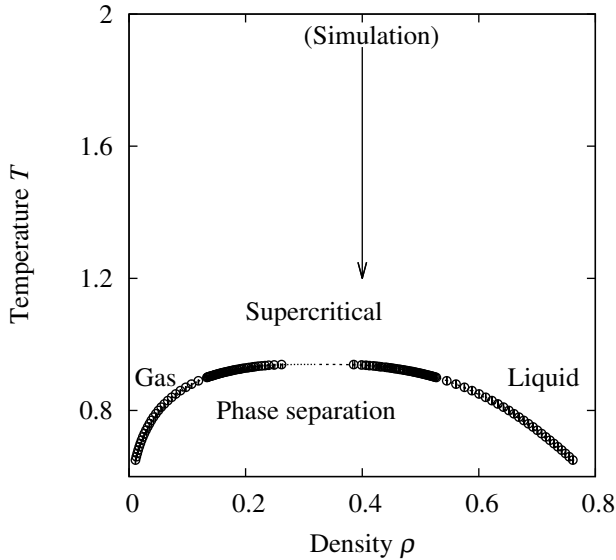}
\caption{Phase diagram of the LJ fluid.
The circles represent the boundaries of liquid-gas coexistence.
The arrow indicates the thermodynamic conditions of the inflow fluid calculated in the present study ($\rho=0.4$).}
\label{f1}
\end{figure}
\begin{figure}[htbp]
\includegraphics{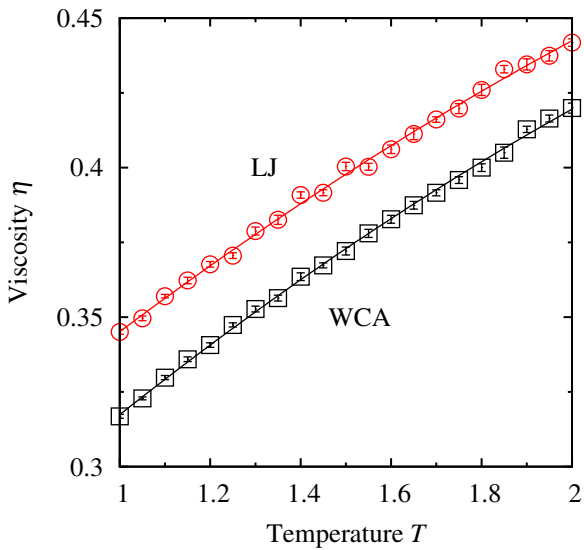}
\caption{Temperature dependence of the viscosity $\eta$ at $\rho=0.4$ for the LJ and WCA fluids.}
\label{f2}
\end{figure}
At the initial state, the fluid particles are randomly distributed with zero overlap within the simulation box.
The initial velocities of fluid particles are given by the Maxwell velocity distribution with an average of 1 in the $x$-direction.
The temperature $T$ of the inflowing fluid is varied in the range of $1\le T \le 2$ at the particle density $\rho=0.4$ (see Fig.~\ref{f1}).
The liquid-gas coexisting envelope was obtained by two-phase simulation~\cite{wih12}.

To compare the difference between single-phase flow and cavitation (LJ) flow,
we used the Weeks--Chandler--Andersen (WCA) fluid~\cite{wca71} (LJ with a cutoff at $r_{\rm c}=2^{1/6}$) for a single-phase flow.
Previous MD simulations of K\'arm\'an vortices used WCA fluids~\cite{awn18,awn19,rc86,rapaport87}.

Numerical integration is performed up to $10\ 000\ 000$ steps with a time step of 0.004 using LAMMPS~\cite{plimpton95}. 
We adopt the velocity-Verlet algorithm for time integration.
The error bars are estimated from three or more independent runs.

To evaluate $Re$, we estimate the viscosities of the LJ and WCA fluids by generating the Poiseuille flow in the MD simulations.
The flow direction is the $x$-direction, and the gravitational acceleration $g=0.0001$ is imposed in the $x$-direction.
The system is a rectangular parallelepiped with a dimension of $L_x\times L_y \times L_z = 100\times 100\times 120$.
The wall is modeled by the Langevin thermostat, which is used for the region of $100 \le z \le 120$.
The viscosity coefficient $\eta$ is obtained by fitting the flow velocity distribution $v_x(z)$ in the $x$-direction in the steady-state to the following equation:
\begin{eqnarray}
v_x(z)&=&\frac{g\rho}{2\eta}z(100-z).
\end{eqnarray}
Figure~\ref{f2} shows the temperature dependence of the viscosity $\eta$.
$Re$ increases with decreasing temperature:
at $T=2$, $1.5$, $1.25$, and $1$, 
$Re=90$, $101$, $107$, and $116$ for the LJ fluid 
and $Re=95$, $107$, $116$, and $126$ for the WCA fluid, respectively.
%
% Results
%
%
\section{Results}
%
%   Non-cavitation flow
%
\subsection{Non-cavitation flow}\label{sec:ncflow}
We first examined the vorticity and density fields for the non-cavitation flow of the LJ fluid at $T=2$ 
(see Figs.~\ref{f3}(a) and \ref{f4}(a) and the top movie in the supplementary material).
To evaluate the vorticity and density fields, we divide the computational domain into cubes with a linear length of $10$.
The vorticity of each cube is estimated by the following formula:
\begin{eqnarray}
\omega_{z} &=& \frac{\partial v_y}{\partial x} - \frac{\partial v_x}{\partial y}, \label{eq3}
\end{eqnarray}
where  $v_\alpha$ is the $\alpha$ component of the flow velocity of each cube.
The derivatives in Eq.~(\ref{eq3}) are calculated using the central difference formula.

\begin{figure}[thbp]
\includegraphics{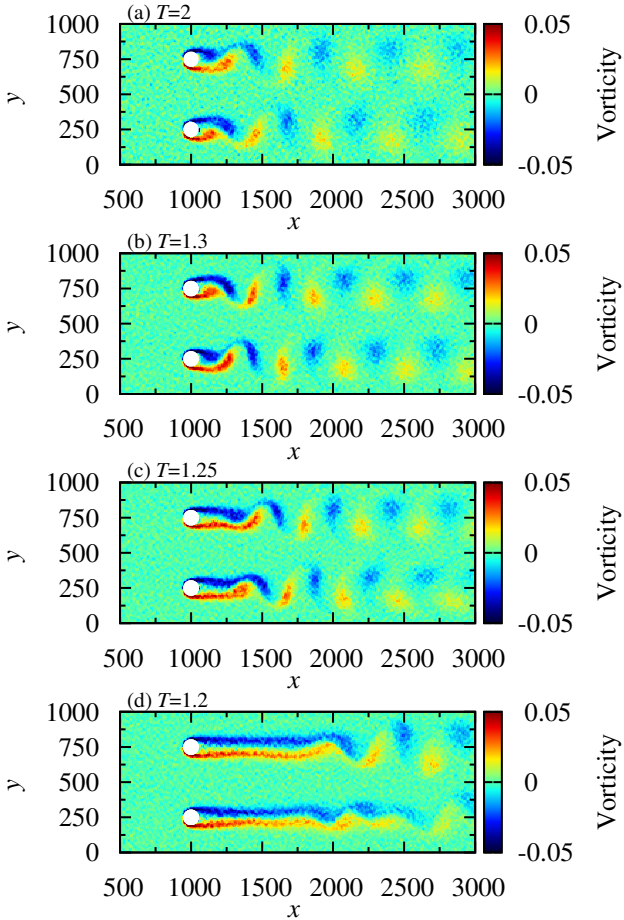}
\caption{Instantaneous vorticity plot of the LJ fluid for (a) non-cavitation flow at temperature $T=2$, and cavitation flow at (b) $T=1.3$, (c) $T=1.25$, and (d) $T=1.2$.}
\label{f3}
\end{figure}

\begin{figure}[thbp]
\includegraphics{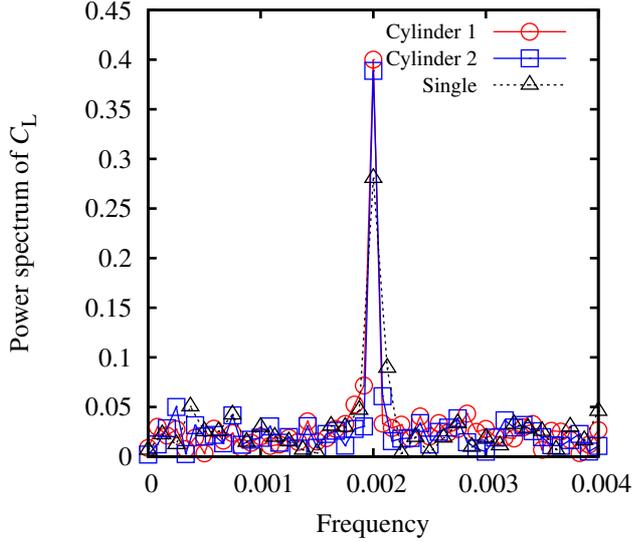}
\caption{Power spectra of the lift coefficient for both cylinders at temperature $T=2$. Dotted line: single-cylinder with the periodic boundary condition.}
\label{f5}
\end{figure}

\begin{figure}[htbp]
\includegraphics{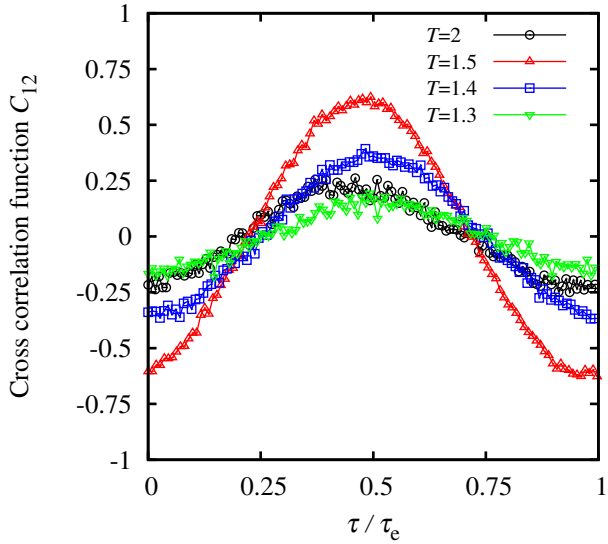}
\caption{Cross-correlation coefficient $C_{12}$ between the lift forces of two cylinders. $\tau_{\rm e}$ is the period of the lift oscillation.}
\label{f6}
\end{figure}

During non-cavitation flow in the LJ fluid, K\'arm\'an vortex streets appear behind the cylinders, synchronizing in the antiphase (Fig.~\ref{f3}(a)).
The density behind the cylinder and the center of the vortex is lower than in other regions (Fig.~\ref{f4}(a)).
To estimate the vortex shedding frequency, the lift coefficient $C_{{\rm L}i}=2F_{{\rm L}i}/(\rho D L_z V^2)$ is Fourier transformed,
where $F_{{\rm L}i}$ is the lift force acting on the $i$th cylinder.
The power spectra of the two cylinders have the same peak position at the frequency $f=0.0022$ (Fig.~\ref{f5}) and the K\'arm\'an vortex shedding from the two cylinders is synchronized.
The phase difference of the vortex shedding between two cylinders is estimated from the cross-correlation coefficient $C_{12}$ of the lift coefficients:
\begin{eqnarray}
C_{12}(\tau)&=&\frac{\left\langle C_{{\rm L}1}\left(t\right)C_{{\rm L}2}\left(t+\tau\right) \right\rangle}{\sqrt{\left\langle C_{{\rm L}1}\left(t\right)^2\right\rangle \left\langle C_{{\rm L}2}\left(t+\tau\right)^2\right\rangle}},
\end{eqnarray}
where $\langle X(t) \rangle$ represents the time average of a function of time $X(t)$.
The cross-correlation coefficient $C_{12}$ at $T=2$ exhibits a clear peak at the phase difference $\pi$ (Fig.~\ref{f6}).
Hence, at a distance between the cylinders of $d=5D$, the vortex shedding synchronizes in the antiphase corroborating previous findings~\cite{pg96}.

The single-cylinder system with the same neighbor distance under periodic boundary condition demonstrates the presence of fluid interactions in the two-cylinder system (Fig.~\ref{f5}).
In the single-cylinder system, it is forced to oscillate in the inphase by the periodic boundary condition.
The size of the simulation box is $L_x \times L_y \times L_z = 3600\times 500\times 100$, and the position of the center axis of the cylinder is $(x, y)=(1000,250)$.
Although the vortex shedding frequency is the same as that of the double-cylinder system, the amplitude of the single-cylinder system is lower than that of the double-cylinder system.
Thus, it is found that the antiphase synchronization enhances the K\'arm\'an vortex oscillation.
\begin{figure}[thbp]
\includegraphics{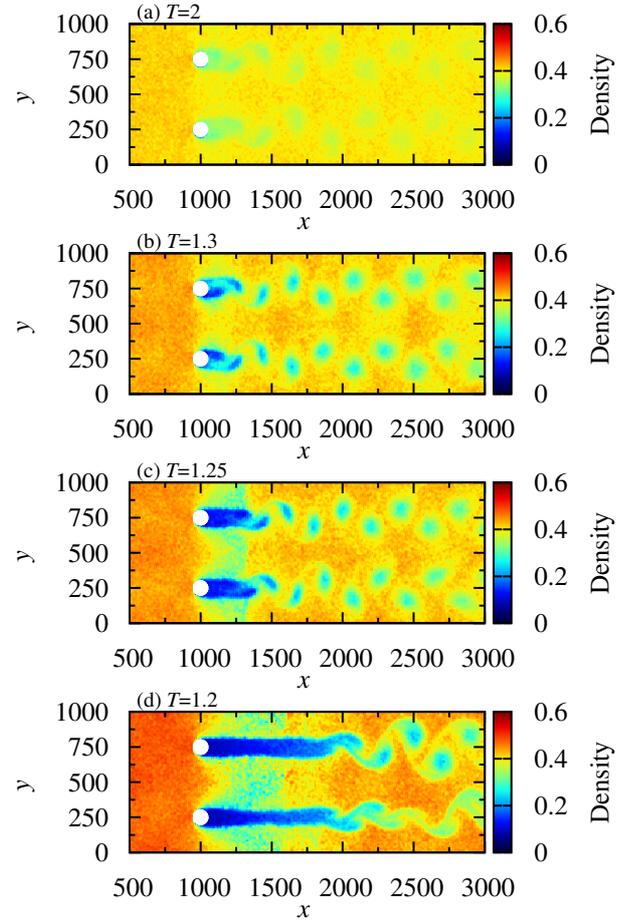}
\caption{Instantaneous density plot of the LJ fluid for (a) non-cavitation flow at temperature $T=2$, and cavitation flow at (b) $T=1.3$, (c) $T=1.25$, and (d) $T=1.2$.}
\label{f4}
\end{figure}
%
% Cavitation flow
%
\subsection{Cavitation flow}\label{sec:cavflow}

Vorticity fields for cavitation flow are shown in Figs.~\ref{f3}(b)--(d) and the bottom three movies in the supplementary material.
The corresponding density fields are shown in Figs.~\ref{f4}(b)--(d), respectively.
As the temperature $T$ decreases, the vortex structure significantly changes due to the density fluctuation induced by the phase transition. 
At $T=1.3$, although K\'arm\'an vortex streets are synchronized in the antiphase as in the case of the non-cavitation flow, the vortex formation length is lengthened.
As shown in Fig.~\ref{f4}(b), because the low-density fluid generated near the cylinder is entrained into the vortex, the vortex formation is delayed. 

At $T=1.25$, the vortex formation length extends further, and the phase of the upper and lower K\'arm\'an vortex streets is slightly disturbed.
A low-density fluid is attached directly behind the cylinder to form a cavity. Vortices are shed from the trailing edge of the cavity (Fig.~\ref{f4}(c)).
At $T=1.2$, the vortex formation length increases remarkably, and the structure of the K\'arm\'an vortex becomes asymmetric (Figs.~\ref{f3}(d) and \ref{f4}(d)).
This asymmetric structure switches with an extended period.

\begin{figure}[thbp]
\includegraphics{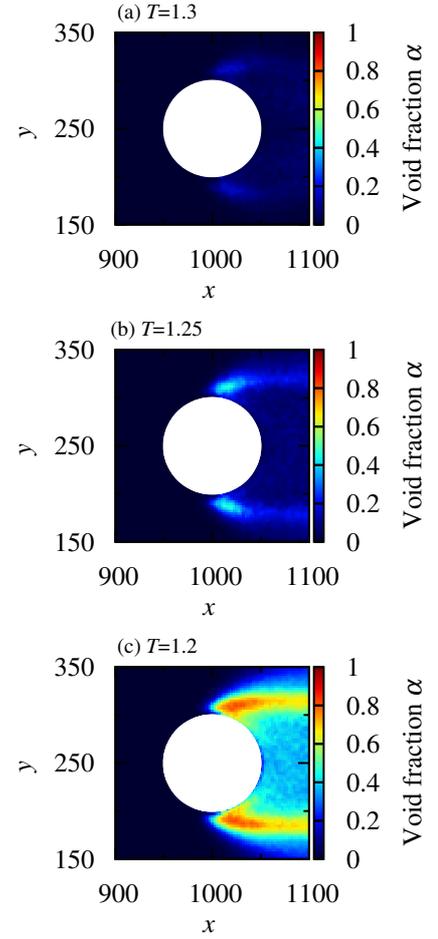}
\caption{Time averaged void fraction near cylinder 1 (white circle) for (a) $T=1.3$, (b) $T=1.25$, and (c) $T=1.2$.}
\label{f7}
\end{figure}
To clarify the effects of bubble generation, we estimate the void fraction near the cylinder under the local equilibrium assumption.
The density in the coexistence region in Fig.~\ref{f1} is expressed as $\rho_{\rm coex}=\rho_{\rm gas}\alpha + \rho_{\rm liq}(1-\alpha)$, where $\rho_{\rm gas}$ and $\rho_{\rm liq}$ are the coexisting densities of the gas and liquid phases, respectively.
Note that the lowest-density region next to the cylinder is not a vacuum but dilute gas (Figs.~\ref{f4}(c) and \ref{f4}(d)).
The time-averaged void fraction $\alpha$ near the cylinder demonstrate that bubbles appear in the vicinity of the separation point,
and the amount of bubble generation increases as the temperature decreases (see Fig.~\ref{f7}).
Therefore, this simulation shows that bubbles significantly affect the vortex shedding cycle and vortex structure.

\begin{figure}[htbp]
\includegraphics{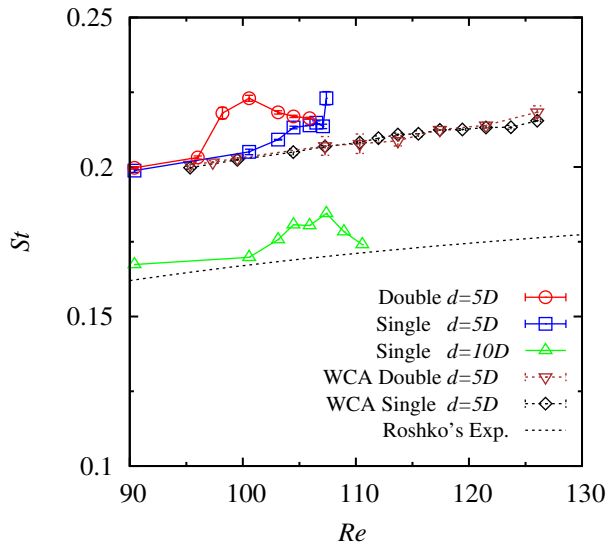}
\caption{Strouhal number $St$ as a function of the Reynolds number $Re$.
Solid lines: LJ fluid; dashed lines: WCA fluid. The broken line denotes Roshko's experimental result given by Eq.~(\ref{f0}).}
\label{f9}
\end{figure}

The generation of bubbles affects the dimensionless vortex shedding frequency $St$.
Figure~\ref{f9} shows the $Re$ dependence of $St$.
The broken line in Fig.~\ref{f9} shows Eq.~(\ref{eq0}).
The lower the temperature, the higher the $Re$ (see Fig.~\ref{f2}).
In the non-cavitation flow condition ($Re\simeq 90$ for the LJ fluid and the whole range of $Re$ for the WCA fluid),
the behavior of $St$ across $Re$ is almost identical between the single- and double-cylinder systems.
In the absence of bubbles, no significant differences exist between the $St$ of the LJ fluid and $St$ of the WCA fluid.
Although $St$ deviates from Eq.~(\ref{eq0}), this result is due to the influence of the neighboring K\'arm\'an vortices.
When $L_y$ is $10D$, the deviation is highly reduced~\cite{awn19}.

As $Re$ increases, the $St$ for the LJ fluid deviates from that of the WCA fluid (Fig.~\ref{f9}).
The LJ fluids exhibit peak $St$ at $Re=101$ and $108$ for the double-cylinder and single-cylinder systems, respectively.
The lower $Re$ peak of the double-cylinder system is presumably due to the increase in vibration by phase entrainment.
The peak height of $C_{12}$ has a maximum at $T=1.5$ that corresponds to the peak position of $St$ (see Fig.~\ref{f6}).
At $Re\gtrsim 110$, the $St$ cannot be measured based on the lift force oscillation of the LJ fluid because no peak appears in the power spectrum.
\begin{figure}[htbp]
\includegraphics{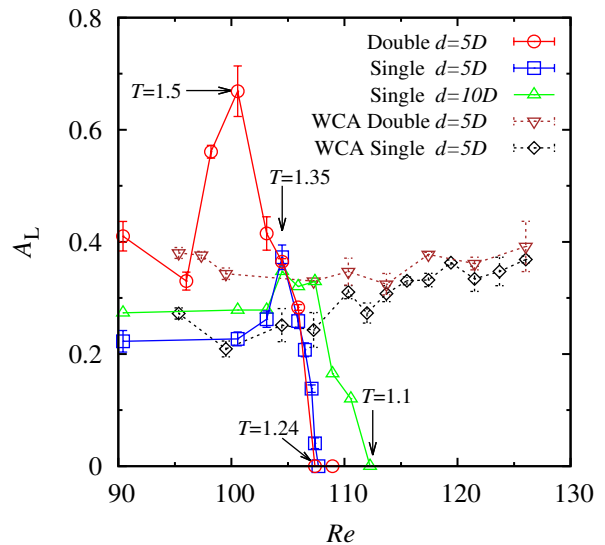}
\caption{Amplitude $A_{\rm L}$ of the lift coefficient $C_{\rm L}$ as a function of the Reynolds number $Re$.
Solid lines: LJ fluid; dashed lines: WCA fluid.}
\label{f10}
\end{figure}
\begin{figure}[htbp]
\includegraphics{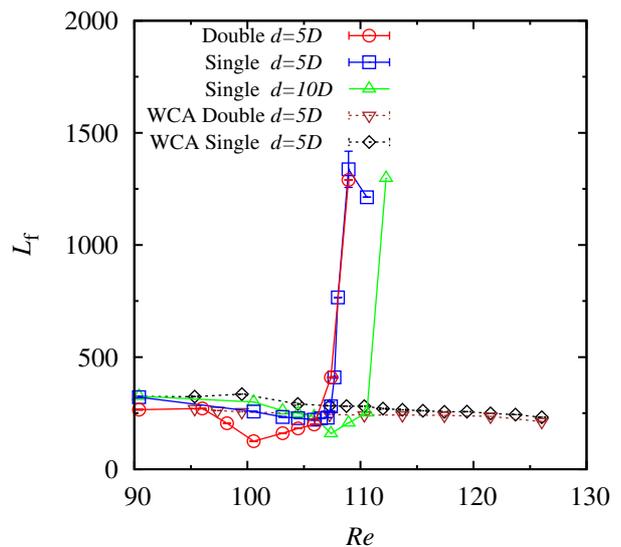}
\caption{Vortex formation length $L_{\rm f}$ as a function of the Reynolds number $Re$.
Solid lines: LJ fluid; dashed lines: WCA fluid.}
\label{f11}
\end{figure}
Figures~\ref{f10} and \ref{f11} show the $Re$ dependence of the lift coefficient amplitude $A_{\rm L}$ and the vortex formation length $L_{\rm f}$, respectively.
We estimate the length $L_{\rm f}$ from a distance between the center axis of the cylinder and the maximum oscillation point of the velocity in $x$-direction behind the cylinder.
In the case of the WCA fluid, $A_{\rm L}$ and $L_{\rm f}$ changes monotonously.
As $Re$ increases, the former increases, and the latter decreases.

For the LJ fluid, the $A_{\rm L}$ and the $L_{\rm f}$ show nonmonotonic behavior due to bubble generation.
Accompanied by the peaks in $St$, the $A_{\rm L}$ and the $L_{\rm f}$ exhibit the maximum and minimum, respectively.
After bubble generation, $A_{\rm L}$ decreases and eventually disappears while $L_{\rm f}$ increases significantly.
This drastic change is caused by bubbles since the bubbles suppress the vortex formation.

When the distance between the cylinders is $10D$,  the $A_{\rm L}$ begins to decrease at higher $Re$ (Fig.~\ref{f10}).
The pressure reduction around the cylinder becomes smaller at the wider channel so that higher $Re$ is required for cavitation.
As cavitation occurs, the lift oscillation eventually disappears in both cases.

In the case of the WCA fluid, $St$, $A_{\rm L}$, and $L_{\rm f}$ for the single-cylinder system are almost the same as for the double-cylinder system.
However, in the case of LJ fluid, the differences appear before bubbles appear, indicating that a precursor appears immediately before the generation of bubbles.
The trends of $St$, the $A_{\rm L}$, and the $L_{\rm f}$ are different between the single-cylinder system and the double-cylinder system until the bubble generation, at which point they become similar.
The bubbles weaken the influence of vortex interference.
Therefore, when the influence of bubbles becomes significant, the vortex shedding characteristics are no longer depend on the phase difference of the K\'arm\'an vortex streets.
Hence, the correlation is reduced by the bubble generation.
%
%
% Discussion
%
%
\section{Discussion}\label{sec:dis}

We found that the vibration of the lift force acting on the cylinder disappears during the cavitation,
whereas the K\'arm\'an vortex still exists, although it begins at a position further from the cylinder.
In the single-cylinder system, the vortex formation length decreases until just before the rapid increase, while the amplitude of the lift coefficient decreases (Figs.~\ref{f10} and \ref{f11}). 
Therefore, the bubbles in the vortex formation region inhibit the propagation of the oscillation induced by the vortex shedding.
The bubbles play an important role in controlling the flow-induced vibration~\cite{zth19} and the drag~\cite{ceccio09} acting on a body.
Previous researchers have concluded that injecting gas into a low-pressure region suppresses vibration and drag~\cite{ztz19}.
In our simulation, the vibration is removed at the cylinder.

In this study, cavitation is induced by change in the local temperature that is accompanied by local pressure change.
Previous studies used pressure change to induce the cavitation.
They reported that $St$ changes with the development of cavitation around a cylinder.
Although $St$ decreases in the cyclic cavity state at which the bubbles appear in conjunction with the vortices, the attached cavity state shows the opposite trend~\cite{fry84}.
In this study, $St$ peaks before the bubble generation starts in the double-cylinder system (Fig.~\ref{f9}).
On the other hand, $St$ increases until the vibration disappears in the case of the single-cylinder system. 
It seems that the left half of the peak is removed by the cavitation.
The amplitude $A_{\rm L}$ and vortex formation length $L_{\rm f}$ also have maximum and minimum at these peak positions.
These precursor peaks are likely due to the formation of invisibly small bubbles or large density fluctuation in the vicinity of the critical point.

The change in the vortex structure shown in this study (Figs.~\ref{f3}(d) and \ref{f4}(d)) is similar results of to Alam {\it et al.}~\cite{azh17}
Those authors examined flow around four cylinders performed by in a single-phase flow.
The behavior of the fluid around the two inner cylinders of the system showed that as the gap between the cylinders became small, two vortex streets in the antiphase changed to an asymmetric structure that changed irregularly over time.
These results indicate that the bubble modifies the effective cylinder diameter to a greater value where the asymmetric wake mode appears.
%
%
% Summary
%
%
\section{Conclusion}\label{sec:sum}

We performed MD simulations to investigate the effects of cavitation on the flow around a circular-cylinder array.
In the non-cavitation flow at high temperatures, the K\'arm\'an vortex street behind the cylinder synchronizes in the antiphase.
The temperature reduction induces the cavitation, since the thermodynamic state of the fluid changes near the cylinder.
The cavitation significantly affects the vortex structure and lift force acting on the cylinder. 
Notably, the vibration of the lift acting on the cylinder decreases and eventually disappears due to the vapor stack between the cylinder and K\'arm\'an vortex.

We also investigated the synchronization of vortices by comparing single-cylinder and double-cylinder systems.
The antiphase synchronization in double-cylinder systems enhances the oscillation of the K\'arm\'an vortices.
Although the cavitation generation point is not changed by this synchronization,
the flow behavior in double-cylinder systems is still altered by neighboring vortex streets after the cavitation.
The asymmetric wake mode was induced by an increase in the effective cylinder diameter due to cavitation, demonstrating that the bubble-induced flow can change the interaction between the neighboring vortex streets.

Unlike simulations based on the Navier-Stokes equations, MD simulations can discuss phase transitions in the flow field on the molecular scale.
Our MD simulation revealed that cavitation occurs in a supercritical fluid without bubble nuclei due to the fluid state change as fluid travels around a cylinder.
MD is a promising tool to investigate cavitation further in various situations such as screw propellers and wings.

\section*{Suplementary material}
See supplementary material for the movies of the time evolution of the vorticity field.

\begin{acknowledgments}
%\acknowledgment
We would like to thank T. Kawakatsu, Y. Morii, S. Tsuda, and Y. Kunishima for helpful discussions.
This research was supported by MEXT as ``Exploratory Challenge on Post-K computer” (Challenge of Basic Science—Exploring Extremes through Multi-Physics and Multi-Scale Simulations) and JSPS KAKENHI Grant No.~JP15K05201.
Computation was partially carried out by using the facilities of the Supercomputer Center, Institute for Solid State Physics, University of Tokyo.
\end{acknowledgments}

%\bibliographystyle{apsrev4-1}
%\bibliography{ref}


\end{document}